\documentstyle[aps,prb,prabib,epsf,multicol]{revtex}
\begin{document}
\draft
\title{Resonant hyper-Raman scattering in spherical quantum dots}
\author{E. Men\'{e}ndez-Proupin and C. Trallero-Giner}
\address{Departamento de F\'\i sica Te\'orica, Universidad de La Habana, 
Vedado \\
10400, La Habana, Cuba}
\author{A. Garc\'\i a-Cristobal}
\address{Departamento de F\'\i sica Aplicada, Universidad de Valencia\\
E-46100 Burjassot, Spain}
\date{\today}
\maketitle

\begin{abstract}
A theoretical model of resonant hyper-Raman scattering by an ensemble of
spherical semiconductor quantum dots has been developed. The electronic
intermediate states are described as Wannier-Mott excitons in the framework
of the envelope function approximation. The optical polar vibrational modes
of the nanocrystallites (vibrons) and their interaction with the electronic
system are analized with the help of a continuum model satisfying both the
mechanical and electrostatic matching conditions at the interface. An
explicit expression for the hyper-Raman scattering efficiency is derived,
which is valid for incident two-photon energy close to the exciton
resonances. The dipole selection rules for optical transitions and
Fr\"ohlich-like exciton-lattice interaction are derived: It is shown that
only exciton states with total angular momentum $L=0,1$ and vibrational
modes with angular momentum $l_p=1$ contribute to the hyper-Raman scattering
process. The associated exciton energies, wavefunctions, and vibron
frequencies have been obtained for spherical CdSe zincblende-type
nanocrystals, and the corresponding hyper-Raman scattering spectrum and
resonance profile are calculated. Their dependence on the dot radius and the
influence of the size distribution on them are also discussed.
\end{abstract}


%\narrowtext	

\begin{multicols}{2}
\section{Introduction}

The research on semiconductor quantum dots (QD's) has undergone a dramatic
increase in recent years, stimulated by their foreseen applications in
optics and electronics technology and also due to their nonlinear optical
properties.\cite{someones} Quantum dot systems based on III-V materials as
well as nanocrystallites of II-VI compounds embedded in glass have been
thoroughly investigated (for a review see Ref.~\onlinecite{2a}). Among the
scattering mechanisms present in polar semiconductors, the optical phonon
emission is known to play a dominant role in QD's, which can be
experimentally investigated by employing a number of methods, such as
infrared absorption and Raman scattering.\cite{A,B,C,D,E,F,G} The
successful interpretation of light scattering by optical phonons relies upon
a good knowledge of the normal vibrational modes. In recent years a
phenomenological continuum theory of optical phonons in nanostructures has
been elaborated\cite{tr3,tr4,tr6}, which is in good agreement with {\it ab
initio} calculations %of the normal modes 
and allows to explain the resonant Raman scattering intensities of phonon
modes induced by interface roughness in quantum wells.\cite{9ch} The theory
has also been generalized to deal with quantum wires and quantum dots, and
used to study resonant Raman scattering in these 
systems.\cite{roca,chamb,eduardo} In particular, 
the formalism is applicable to II-VI
semiconductor nanocrystallites embedded in glass since they can have
dimensions as small as 13~\AA {}, in which case the mechanical confinement 
of optical
vibrational modes has strong effects (in a QD the concept of phonon as an
excitation in a periodic system labeled by a wave vector is lost, and
therefore we will use the term {\it vibron} to denote the QD vibrational
modes). As it has been theoretically demonstrated, optical vibrons in
spherical nanocrystallites have a mixed longitudinal-transverse character
and can exhibit different symmetries.\cite{roca} Raman scattering provides a
useful tool to investigate experimentally these vibrons, but only
spherically symmetric modes are accessible to this technique. In the search
of complementary experimental techniques which can overcome this limitation,
hyper-Raman (HR) spectroscopy appears to be a suitable candidate.\cite
{Alberto2} Recently, HR spectroscopy has been used to study the optical
vibrational modes of CdS and CuBr QD's.\cite{H,I,J} In this paper we present
a theoretical model which allows to study hyper-Raman scattering (HRS) by
optical vibrons under resonance conditions, and illustrate it by performing
numerical calculations of the scattered intensities in CdSe
nanocrystallites. As far as the exciton-lattice interaction is concerned
only the Fr\"ohlich coupling is considered here, though it is admitted that
the deformation potential interaction might be of importance for photon
energies far from the excitonic resonances. The selection rules for the
scattering process are worked out and it is shown that in fact HRS can be
used to probe optical vibrations with non-spherical symmetry.

The paper is organized as follows. The main concepts underlying the
hyper-Raman scattering and the description of the nanocrystal vibrational
modes and excitons are outlined in Sec.~\ref{II}. Section~\ref{III} contains
the theoretical expressions for the matrix elements and HRS efficiency,
which are used to analyze the selection rules of the scattering process. The
calculation of the exciton and vibrational spectra of CdSe nanocrystallites
are presented in Sec.~\ref{IV}, and the numerical results obtained for the
HRS efficiency are reported and discussed in Sec.~\ref{V}. Section~\ref{VI}
is devoted to the main conclusions of the work and final comments.

\section{Theory}

\label{II}

\subsection{Hyper-Raman efficiency}

Hyper-Raman scattering is a nonlinear process which consists of the
absorption of two photons of frequency $\omega _i$, wave vector ${\bf k}_i$,
and polarization ${\bf e}_i$ ($i=1,2$), and the emission of one photon ($%
\omega _s$,${\bf k}_s$,${\bf e}_s$) with the simultaneous excitation of a
number of vibrational modes.\cite{levenson} Due to its nonlinear nature, it
is convenient to express the HR yield by the normalized
(intensity-independent) scattering cross section\cite{12a} 
\begin{equation}
\frac 1{I_i}\frac{d^{\,2}\sigma ^{ij}}{d\omega _sd\Omega _s}=\frac 1{I_iI_j}%
\frac{d^{\,2}P_{ij}}{d\omega _sd\Omega _s}\quad ,  \label{ec:1}
\end{equation}
$I_i$ and $I_j$ being the excitation intensities and $\frac{d^{\,2}P_{ij}}{%
d\omega _sd\Omega _s}$ the scattered power per unit of solid angle $\Omega
_s $ and unit frequency. The %normalized differential cross section or
scattered power can be related to the probability per unit time $W(\omega
_i,\omega _j,\omega _s;R)$ of a single scattering event in a QD of radius $R$%
, so that we can express the normalized differential cross section as\cite
{12a} 
\begin{equation}
\frac 1{I_i}\frac{d^{\,2}\sigma ^{ij}}{d\omega _sd\Omega _s}=\frac{V^3}{%
(2\pi )^3}\, \frac{\omega _s^3}{\omega _i}\frac 1{\hbar \omega _j}\frac{\eta
_i\eta _j\eta _s^3}{c^5}\,W(\omega _i,\omega _j,\omega _s;R) \quad ,
\end{equation}
where $V$ is the normalization volume, $\eta _i$ ($i=1,2$) and $\eta _s$ are
the nanocrystal refraction indices at $\omega _i$ and $\omega _s$,
respectively, and $c$ is the velocity of light in vacuum. According to the
Fermi's Golden Rule, $W$ can be written as 
\begin{eqnarray}
W(\omega _i,\omega _j,\omega _s;R)=\frac{2\pi }{\hbar ^2}\sum_p|M_p(\omega
_i,\omega _j,\omega _s;R)|^2
\nonumber \\
\times\delta (\omega _i+\omega _j-\omega _p-\omega
_s)\quad ,  \label{ec:3}
\end{eqnarray}
where $M_p$ is the scattering amplitude. Each scattering event is
accompanied by the emission of a vibron of frequency $\omega _p$, and the
total scattering probability is obtained after summing over all possible
vibron states $p$. Only vibron emission (Stokes) processes will be
considered in this work.

The delta function in (\ref{ec:3}) can be eventually replaced by a
Lorentzian, 
\begin{eqnarray}
\delta(\omega_i+\omega_j-\omega_p-\omega_s) \longrightarrow 
\nonumber \\
\Delta(\omega_s)=%
\frac{\hbar}{\pi} \frac{\Gamma_p} {(\hbar\omega_i+\hbar\omega_j-\hbar%
\omega_p-\hbar\omega_s)^2 +\Gamma_p^2} \quad ,
\end{eqnarray}
in order to take into account the effect of the vibron lifetime $%
\tau_p=\hbar/\Gamma_p$ on the line broadening of the HR spectrum.

The scattering amplitude in (\ref{ec:3}) can be calculated in fourth-order
perturbation theory. Under resonance conditions the important contributions
to $M_p$ are illustrated by the diagrams in Fig.~1. In Fig.~1(a), the
exciton created in the state $|\mu _1\rangle $ after the absorption of a
photon ($\omega _i,{\bf k}_i,{\bf e}_i$) is first scattered to the state $%
|\mu _2\rangle $ by the absorption of a second photon ($\omega _j,{\bf k}_j,%
{\bf e}_j$). In the next step, the interaction with the lattice induces an
excitonic transition from $\left|\mu_2\right>$ to $\left|\mu_3\right>$,
accompanied by the creation of a vibron of frequency $\omega _p$. The
exciton finally recombines emitting a photon ($\omega _s, {\bf k}_s,{\bf e}%
_s $). The amplitude corresponding to this process is
\end{multicols}
\begin{equation}
M_p^{(a)}=\sum_{\mu _1,\mu _2,\mu _3}\frac{\langle F|\hat H_{E-R}^{(s)}|\mu
_3\rangle \langle \mu _3|\hat H_{E-L}|\mu _2\rangle \langle \mu _2|\hat H%
_{E-R}^{(j)}|\mu _1\rangle \langle \mu _1|\hat H_{E-R}^{(i)}|I\rangle }{%
(\hbar \omega _s-E_{\mu _3}+i\Gamma _{\mu _3})(\hbar \omega _i+\hbar \omega
_j-E_{\mu _2}+i\Gamma _{\mu _2})(\hbar \omega _i-E_{\mu _1}+i\Gamma _{\mu
_1})}\quad ,  \label{ec:4}
\end{equation}
\begin{multicols}{2}
where $\left| I\right\rangle $ ($\left| F\right\rangle $) is the initial
(final) state of the scattering process, $E_{\mu _i}$ and $\Gamma _{\mu _i}$
are the energies and lifetime broadenings of the excited electronic states $%
\left|\mu_i\right>$ in the QD. $\hat H_{E-L}$ and $\hat H_{E-R}$ are the
Hamiltonian operators for the interaction of the electronic system with the
lattice and radiation field, respectively.

Since the interesting range of $\hbar \omega _i+\hbar \omega _j$ for
resonant HR spectroscopy occurs around the fundamental absorption edge $E_0$,
the relevant resonances in the matrix element (\ref{ec:4}) will occur at
energies 
\begin{mathletters}
\label{ec:7}
\begin{eqnarray}
\hbar \omega _i+\hbar \omega _j &=&E_{\mu _2}\qquad {\rm (incoming\
resonance)}\quad , \\
\hbar \omega _s &=&E_{\mu _3}\qquad {\rm (outgoing\ resonance)}\quad .
\end{eqnarray}
\end{mathletters}
For the process shown in Fig.~1(b) the scattering amplitude is 
\end{multicols}
\begin{equation}
M_p^{(b)}=\sum_{\mu _1,\mu _2,\mu _3}\frac{\langle F|\hat H_{E-R}^{(s)}|\mu
_3\rangle \langle \mu _3|\hat H_{E-R}^{(j)}|\mu _2\rangle \langle \mu _2|%
\hat H_{E-L}|\mu _1\rangle \langle \mu _1|\hat H_{E-R}^{(i)}|I\rangle }{%
(\hbar \omega _s-E_{\mu _3}+i\Gamma _{\mu _3})(\hbar \omega _i-\hbar \omega
_p-E_{\mu _2}+i\Gamma _{\mu _2})(\hbar \omega _i-E_{\mu _1}+i\Gamma _{\mu
_1})}\quad .  \label{ec:8}
\end{equation}
\begin{multicols}{2}
In Eqs.~(\ref{ec:4}) and (\ref{ec:8}) it is implicitly 
understood that the sum over
non-topologically equivalent diagrams (i$\leftrightarrow $j) must be taken
into account. The examination of the energy denominators in Eq.~(\ref{ec:8})
clearly indicates that $|M_p^{(b)}|\ll |M_p^{(a)}|$ in the resonance region $%
\hbar \omega _i+\hbar \omega _j\sim E_0$. Therefore the contribution from $%
M_p^{(b)}$ has been dropped out from our theoretical model.

Finally, it must be noted that in real samples the nanocrystallites present
a distribution over size and shape. We intend to study here HRS by an
ensemble of spherical QD's characterized by a distribution over radii $F(R)$%
. The corresponding average (intensity-independent) HRS efficiency\cite
{lscatter} is given by 
\begin{equation}  \label{new}
\left\langle \frac{d^{\,2}S_{HR}^{ij}}{d\omega _sd\Omega _s}\right\rangle =%
\frac 1{\langle V_D\rangle }\int \frac 1{I_i}\frac{d^{\,2}\sigma ^{ij}}{%
d\omega _sd\Omega _s}\,F(R)dR\quad ,  \label{ec:8:0}
\end{equation}
where $\langle V_D\rangle $ is the average quantum dot volume.

\subsection{Exciton states}

The intermediate electronic virtual states appearing in the HR process
(\thinspace see Eq.~(\ref{ec:4})\thinspace ) are taken to be size-confined
Wannier-Mott excitons, treated in the framework of the envelope function
approximation (EFA). The details of this approach can be found in Ref.~%
\onlinecite{eduardo}, where it is employed for the study of resonant Raman
scattering in QD's. We summarize here, for notation purposes, the main
expressions of the exciton model. In the absence of electron-hole
interaction, the wave function of an uncorrelated electron-hole pair (EHP)
in a spherical QD can be written as 
\begin{eqnarray}
\Phi _{\alpha ,L,M}({\bf r}_e,{\bf r}_h)=\sum_{m_e,m_h}(l_e\,l_h\,m_e\,m_h|L%
\,M)
\nonumber \\
\phi _{n_e,l_e,m_e}({\bf r}_e)\,\phi _{n_h,l_h,m_h}({\bf r}_h)\quad ,
\label{ec:12}
\end{eqnarray}
where $\alpha $ is an abbreviated notation for $\alpha \equiv
(n_e,n_h,l_e,l_h)$. The single-particle (electron or hole) wave functions
are given by 
\begin{equation}
\phi _{n,l,m}({\bf r})=\,R_{n,l}(r)\,Y_{l,m}(\Omega )\quad .  \label{ec:13}
\end{equation}
$Y_{l,m}(\Omega )$ ($l=0,1,\dots $ and $m=-l,...,l$) are the spherical
harmonics\cite{jackson}, and $R_{n,l}(r)$ is the radial part of the wave
function, which depends on the confinement potential. The corresponding
energy is denoted by $E_{n,l}$. $L$ and $M$ are the quantum numbers
corresponding to the squared total angular momentum, $\hat {{\bf L}}^2$, and
its projection along the $z$-axis, $\hat L_z$, respectively, and $%
(l_e\,l_h\,m_e\,m_h|L\,M)$ are the Clebsch-Gordan coefficients.\cite{brink}

The eigenfunctions of the full Hamiltonian including the electron-hole
Coulomb interaction (exciton wave functions) are expanded in terms of the
EHP states (\ref{ec:12}), 
\begin{eqnarray}  
\Psi_{\mu}({\bf r}_e,{\bf r}_h)\equiv \Psi_{N,L,M,P}({\bf r}_e,{\bf r}_h)=
\nonumber \\
\sum_{\alpha} C_{N,L,M,P}(\alpha)\,\Phi_{\alpha,L,M}({\bf r}_e,{\bf r}_h)
\quad ,
\label{ec:14}
\end{eqnarray}
where $N$ and $P$ are the quantum numbers corresponding to the energy and
inversion operator, respectively. Note that the state defined by Eq.~(\ref
{ec:14}) has definite parity: $P=1$ (even parity) if $l_e+l_h$ is even and $%
P=-1$ (odd parity) if $l_e+l_h$ is odd. By applying the matrix
diagonalization technique described in Ref.~\onlinecite{eduardo} we obtain
the exciton energies $E_{\mu}$ as the eigenvalues of the secular equation
and the coefficients $C_{\mu}(\alpha)$ as the corresponding eigenvectors.

\subsection{Exciton-lattice interaction}

Excitons in semiconductors interact with lattice vibrations by means of
deformation potential and piezoelectric interactions. In polar materials,
the long-range electrostatic field associated to the optical vibrations
introduces a different coupling mechanism, the Fr\"ohlich interaction. It is
now well established that, despite its dipole-forbidden character\cite
{lscatter}, the Fr\"ohlich interaction plays an important role in one-phonon
Raman scattering by bulk zincblende semiconductors.\cite{tr1} On the other
hand, in systems which lack translational invariance, like QD's,
Fr\"ohlich-induced Raman scattering becomes allowed and therefore it is
expected to have a dominant role. Accordingly, in our model for HRS we
consider only the Fr\"ohlich-like interaction between excitons and vibrons.

In order to describe the polar optical vibrational modes (vibrons) of
spherical QD's we rely upon the results of Ref.~\onlinecite{roca}, where a
macroscopic continuum model coupling the mechanical displacement and the
electrostatic potential is developed. For the case of complete confinement
it has been shown that the normal modes with non vanishing electrostatic
potential exhibit a mixed longitudinal-transverse character. They are
labeled by a set of integer numbers $p\equiv (n_p,l_p,m_p)$, which are
related to their symmetry properties. More details on the calculation of the
displacement, electrostatic potential and frequency $\omega_{n_p,l_p}$
associated to these modes can be found in Refs.~\onlinecite{roca} and %
\onlinecite{chamb}.

The exciton-vibron interaction Hamiltonian operator can be written as 
\begin{equation}
\hat H_{E-L}=e\,\hat \varphi _F({\bf r}_e)-e\,\hat \varphi _F({\bf r}%
_h)\quad ,  \label{ec:18}
\end{equation}
where $-e$ ($e>0$) is the electron charge and 
\begin{eqnarray}
e\,\hat \varphi _F({\bf r})=\frac{C_F}{\sqrt{R}}\sum_{n_p,l_p,m_p}\sqrt{%
\frac{\omega _L}{\omega _{n_p,l_p}}}\,\Phi _{n_p,l_p}(r)
\nonumber \\
\times\left(
\,Y_{l_p,m_p}(\Omega )\,\hat b_{n_p,l_p,m_p}+H.c.\,\right) \quad .
\label{ec:19}
\end{eqnarray}
Here $\hat b_{n_p,l_p,m_p}$ is the vibron annihilation operator and $H.c.$
means Hermitian conjugate. The explicit form of the radial function $\Phi
_{n_p,l_p}(r)$ can be found in Refs.~\onlinecite{roca,chamb}. Since $\omega
_{n_p,l_p}\sim \omega _L$ ($\omega _L$ is the bulk LO-phonon frequency at
the $\Gamma $-point), we will omit hereafter the factor $\sqrt{\omega
_L/\omega _{n_p,l_p}}$ in (\ref{ec:19}). The Fr\"ohlich constant $C_F$ is given by 
\begin{equation}
C_F=\sqrt{2\pi \,e^2\,\hbar \omega _L\,\left( \frac 1{\epsilon _{a\infty }}-%
\frac 1{\epsilon _{a0}}\right) }\quad ,  \label{CF}
\end{equation}
$\epsilon _{a\infty }$ and $\epsilon _{a0}$ being the high- and
low-frequency dielectric constants of the crystallite material.

\section{Matrix elements and selection rules}

\label{III}

In this section we analyze in detail the matrix elements appearing in Eq.~(%
\ref{ec:4}), and derive from them the selection rules for the exciton and
vibron states which participate in the HR process.

The matrix elements $\langle \mu _1|\hat H_{E-R}^{(i)}|I\rangle $ and $%
\langle F|\hat H_{E-R}^{(s)}|\mu _3\rangle $) for direct allowed optical
transitions between valence ($v$) and conduction ($c$) bands are given by%
\cite{eduardo} 
\begin{equation}
\langle \mu _1|\hat H_{E-R}^{(i)}|I\rangle =\displaystyle \frac e{m_0}\frac 1%
{\sqrt{V}}\sqrt{\frac{2\pi \hbar }{\omega _i\eta _i^2}}\,{({\bf e}_i\cdot 
{\bf p}_{cv})}\,f_{\mu _1}\quad ,  \label{ec:23}
\end{equation}
where ${\bf p}_{cv}$ is the interband momentum matrix element between
valence and conduction Bloch functions at ${\bf k}=0$, and $m_0$ is the
free-electron mass. A similar expression can be obtained for $\langle \mu _3|%
\hat H_{E-R}^{(s)}|F\rangle $. The exciton overlap integral $f_\mu $ is
given by 
\begin{eqnarray}
f_\mu \equiv f_{N,L,M,P}=\int \Psi _{N,L,M,P}({\bf r},{\bf r})\,d^3{\bf r} 
=\delta _{L,0}\,\delta _{M,0}\,
\nonumber \\
\times\delta_{P,1}\sum_{n_e,n_h}\,\sum_l(-1)^l\,%
\sqrt{2l+1}\,C_{N,0,0,1}(n_e,n_h,l,l)
\nonumber \\
\int_0^\infty
R_{n_e,l}(r)\,R_{n_h,l}(r)\,r^2\,dr\quad .  \label{ec:24}
\end{eqnarray}

Hence, the annihilation of the first incoming photon with frequency $\omega
_i$ creates an exciton in the state with zero angular momentum and even
parity, 
\[
\mu _1\equiv (N_1,L_1=0,M_1=0,P_1=1)\quad , 
\]
and analogously, the scattered photon of frequency $\omega _s$ is emitted
upon the recombination of an exciton in the state 
\[
\mu _3\equiv (N_3,L_3=0,M_3=0,P_3=1)\quad . 
\]
Both exciton states have $L=0$ and $P=1$ because the interband transitions
induced by the radiation field require the electron and hole to have equal
orbital angular momentum quantum numbers, $l_e=l_h$.

A different situation is found when considering the scattering between
exciton states induced by the second incoming photon (\thinspace matrix
element $\langle \mu _2|\hat H_{E-R}^{(j)}|\mu _1\rangle $ in Eq.~(\ref{ec:4}%
)\thinspace ). Two possibilities arise in this case \cite{Alberto3}: On the
one hand there may be transitions accompanied by scattering of the electron
(or hole) between QD levels corresponding to different bands ({\it interband
excitonic transitions}), and by the other hand there are also transitions in
which the electron and hole remain in the same band, and only the exciton
envelope function is affected by the interaction with the radiation field (%
{\it intraband excitonic transitions}). In III-V and II-VI compounds, and
for $\hbar \omega _i+\hbar \omega _j$ around the fundamental absorption
edge, the interband excitonic transitions are due to remote bands which are
far enough in energy, thus giving a negligible contribution to the
scattering amplitude $M_p^{(a)}$. Therefore we consider only intraband
excitonic transitions, whose matrix element can be written, in the dipole
approximation,\cite{Alberto3} as 
\begin{eqnarray}
\langle \mu _2|\hat H_{E-R}^{(j)}|\mu _1\rangle =\int \,\Psi _{\mu _2}^{*}(%
{\bf r}_e,{\bf r}_h)
\nonumber \\
\left( \hat H_{e-R}^{(j)}-\hat H_{h-R}^{(j)}\right) \Psi
_{\mu _1}({\bf r}_e,{\bf r}_h)\,d^3{\bf r}_e\,d^3{\bf r}_h\quad ,
\label{ec:25}
\end{eqnarray}
where 
\begin{equation}
\hat H_{\nu -R}^{(j)}=\frac e{m_\nu }\frac 1{\sqrt{V}}\sqrt{\frac{2\pi \hbar 
}{\omega _j\eta _j^2}}\,{({\bf e}_j\cdot \hat {{\bf p}}_\nu )}\quad (\nu
=e,h)\quad ,  \label{ec:26}
\end{equation}
$\hat {{\bf p}}_\nu =-i\hbar \,\bbox{\nabla}_\nu $ is the linear momentum
operator, and $m_e$ ($m_h$) is the electron (hole) effective mass (taken to
be positive).

Using the expansion (\ref{ec:14}) in (\ref{ec:25}) we get 

\end{multicols}

\begin{eqnarray}
\langle \mu _2|\hat H_{E-R}^{(j)}|\mu _1\rangle ={e}\,\frac 1{\sqrt{V}}\sqrt{%
\frac{2\pi \hbar }{\omega _j\eta _j^2}}\sum_{\alpha ,\alpha ^{\prime
}}\,C_{N_2,L_2,M_2,P_2}^{*}(\alpha ^{\prime })\,C_{N_1,L_1,M_1,P_1}(\alpha )
\left\langle \alpha ^{\prime },L_2,M_2\left| \frac{{\bf e}_j\cdot 
\hat {{\bf p}}_e}{m_e}-\frac{{\bf e}_j\cdot \hat {{\bf p}}_h}{m_h}\right|
\alpha ,L_1,M_1\right\rangle \quad .  \label{ec:26:1}
\end{eqnarray}
Let us now concentrate on the electron part of the matrix element appearing
in (\ref{ec:26:1}). By making use of the operator identity $\hat {{\bf p}}%
_e=\left( im_e/\hbar \right) \left[ \hat H_e,{\bf r}_e\right] $, where $\hat 
H_e$ is the single-electron Hamiltonian, it is easy to find that 
\begin{eqnarray}
{\left\langle n_e^{\prime },n_h^{\prime },l_e^{\prime },l_h^{\prime
},L_2,M_2\left| \frac{{\bf e}\cdot \hat {{\bf p}}_e}{m_e}\right|
n_e,n_h,l_e,l_h,L_1,M_1\right\rangle =}\,\delta _{n_h^{\prime },n_h}\,\delta
_{l_h^{\prime },l_h}
\nonumber \\
\times \frac i\hbar (E_{n_e^{\prime },l_e^{\prime }}-E_{n_e,l_e}){\langle
n_e^{\prime },n_h^{\prime },l_e^{\prime },l_h^{\prime },L_2,M_2|{\bf e}\cdot 
{\bf r}_e|n_e,n_h,l_e,l_h,L_1,M_1\rangle }\quad .  \label{ec:29}
\end{eqnarray}
\begin{multicols}{2}
The matrix element in the previous equation can be greatly simplified by
taking advantage of the theory of angular momentum, as it is explicitly
shown in the Appendix. Moreover, for $L_1=M_1=0$ ($l_e=l_h$), Eq.~(\ref
{ec:29}) reduces to: 
\end{multicols}
\begin{eqnarray}
{\left\langle n_e^{\prime },n_h^{\prime },l_e^{\prime },l_h^{\prime
},L_2,M_2\left| \frac{{\bf e}\cdot \hat {{\bf p}}_e}{m_e}\right|
n_e,n_h,l_e,l_e,0,0\right\rangle =}\,\delta _{n_h^{\prime },n_h}\,\delta
_{l_h^{\prime },l_h}\,\delta _{L_2,1}\,\delta _{P_2,-1}a_{M_2}
\nonumber \\
\times \frac{i\,R}\hbar \frac 1{\sqrt{3}}(E_{n_e^{\prime },l_e^{\prime
}}-E_{n_e,l_e})\left[ \sqrt{\frac{l_e+1}{2l_e+1}}\delta _{l_e^{\prime
},l_e+1}-\sqrt{\frac{l_e}{2l_e+1}}\delta _{l_e^{\prime },l_e-1}\right]
G_{n_e,l_e\rightarrow n_e^{\prime },l_e^{\prime }}\quad ,  \label{ec:32}
\end{eqnarray}
\begin{equation}
G_{n,l\rightarrow n^{\prime },l^{\prime }}=\frac 1R\int_0^\infty
R_{n^{\prime },l^{\prime }}(r)\,R_{n,l}(r)\,r^3\,dr\quad .  \label{ec:33}
\end{equation}
\begin{multicols}{2}
The final expression for the matrix element (\ref{ec:26:1}) can be arranged
in the form 
\end{multicols}
\begin{equation}
\langle \mu _2|\hat H_{E-R}^{(j)}|\mu _1\rangle =\frac{ie\,E_0\bar R^2}{%
\hbar R}\frac 1{\sqrt{V}}\sqrt{\frac{2\pi \hbar }{\omega _j\eta _j^2}}\left[ 
{\cal F}_{\mu _1\rightarrow \mu _2}^{(e)}-{\cal F}_{\mu _1\rightarrow \mu
_2}^{(h)}\right] \quad ,  \label{ec:34}
\end{equation}
where we have introduced the energy $E_0=\hbar ^2/2m_0\bar R^2$ ($\bar R$ is
the average QD radius) to make ${\cal F}^{(e,h)}$ dimensionless. The
explicit expression for ${\cal F}_{\mu _1\rightarrow \mu _2}^{(e)}$ is 

\begin{eqnarray}
{\cal F}_{\mu _1\rightarrow \mu _2}^{(e)}=\delta _{L_2,1}\,\delta
_{P_2,-1}\,a_{M_2}\sum_{\alpha ,\alpha ^{\prime }}\delta _{l_e,l_h}\delta
_{l_h^{\prime },l_h}\,\delta _{n_h^{\prime },n_h}C_{\mu _2}^{*}(\alpha
^{\prime })\,C_{\mu _1}(\alpha )
\nonumber \\
\times \left( \frac R{\bar R}\right) ^2\frac{(E_{n_e^{\prime },l_e^{\prime
}}-E_{n_e,l_e})}{\sqrt{3}E_0}\left[ \sqrt{\frac{l_e+1}{2l_e+1}}\delta
_{l_e^{\prime },l_e+1}-\sqrt{\frac{l_e}{2l_e+1}}\delta _{l_e^{\prime
},l_e-1}\right] G_{n_e,l_e\rightarrow n_e^{\prime },l_e^{\prime }}\quad .
\label{ec:35}
\end{eqnarray}
\begin{multicols}{2}
A similar expression holds for ${\cal F}_{\mu _1\rightarrow \mu _2}^{(h)}$
after the exchange of the subscripts $e$ and $h$.

An important consequence to be drawn from (\ref{ec:35}) is that after the
absorption of the second incoming photon, the exciton must be in the state: 
\[
\mu _2\equiv (N_2,L_2=1,M_2=0,\pm 1,P_2=-1)\quad .
\]
Otherwise stated, two-photon absorption generate excitons in $L=1$ states,
in contrast to one-photon transitions for which the final exciton state is
necessarily $L=0$. Moreover, the factor $\delta _{l^{\prime },l\pm 1}$ in
Eq.~(\ref{ec:35}) introduce the parity selection rule $P_1=1\to P_2=-1$,
indicating that the intraband transitions are accompanied by a change in the
parity of the excitonic state. If the incident light is linearly polarized
with ${\bf e}_j\parallel \hat z$ then $a_{M_2}=\delta _{M_2,0}$ (see the
Appendix), whereas for circular polarization ${\bf e}_j\parallel (\hat x\pm i%
\hat y)/\sqrt{2}$ we have $a_{M_2}=\delta _{M_2,\pm 1}$ (\thinspace $\hat x,%
\hat y$, and $\hat z$ represent some system of orthogonal axes attached to
the laboratory frame\thinspace ). Hence, the $L_2=1$ excitonic states
participating in the HRS process have $M_2=0$ for linearly polarized light
and $M_2=\pm 1$ for circularly polarized light.

Now we turn to the matrix element of the exciton-lattice interaction $%
\langle \mu _3|\hat H_{E-L}|\mu _2\rangle $. If we select the vibron
creation terms in (\ref{ec:19}) and make use of (\ref{ec:14}), we get the
expression 
\end{multicols}
\begin{eqnarray}
\langle \mu _3|\hat H_{E-L}|\mu _2\rangle =\frac{C_F}{\sqrt{R}}\sum_{\alpha
^{\prime },\alpha ^{\prime \prime }}\,C_{N_3,L_3,M_3,P_3}^{*}(\alpha
^{\prime \prime })C_{N_2,L_2,M_2,P_2}(\alpha ^{\prime })
\nonumber \\
\times \sum_{n_p,l_p,m_p}\,\left\langle \alpha ^{\prime \prime
},L_3,M_3\left| \Phi _{n_p,l_p}(r_e)\,Y_{l_p,m_p}^{*}(\Omega _e)-\Phi
_{n_p,l_p}(r_h)\,Y_{l_p,m_p}^{*}(\Omega _h)\right| \alpha ^{\prime
},L_2,M_2\right\rangle \quad ,  \label{ec:38}
\end{eqnarray}
\begin{multicols}{2}
which consists again of separated electron and hole contributions, each of
them being a sum of matrix elements of spherical tensors. Therefore the
method used previously for the intraband exciton-photon matrix elements (see
the Appendix) applies also here. By following that procedure, and
considering that $L_2=1,L_3=M_3=0$, Eq.~(\ref{ec:38}) is reduced to 
\begin{equation}
\langle \mu _3|\hat H_{E-L}|\mu _2\rangle =\frac{C_F}{\sqrt{R}}\left[ {\cal H%
}_{\mu _2\rightarrow \mu _3}^{(e)}-{\cal H}_{\mu _2\rightarrow \mu
_3}^{(h)}\right] \quad ,  \label{ec:41}
\end{equation}
where 
\end{multicols}
\begin{eqnarray}
{\cal H}_{\mu _2\rightarrow \mu _3}^{(e)}=\delta _{l_p,1}\delta
_{m_p,M_2}\sum_{\alpha ^{\prime },\alpha ^{\prime \prime }}\delta
_{l_e^{\prime \prime },l_h^{\prime \prime }}\,\delta _{n_h^{\prime \prime
},n_h^{\prime }}\,\delta _{l_h^{\prime \prime },l_h^{\prime }}C_{\mu
_3}^{*}(\alpha ^{\prime \prime })\,C_{\mu _2}(\alpha ^{\prime })
\nonumber \\
\times \frac 1{\sqrt{4\pi }}\left[ \sqrt{\frac{l_e^{\prime \prime }+1}{%
2l_e^{\prime \prime }+1}}\delta _{l_e^{\prime },l_e^{\prime \prime }+1}-%
\sqrt{\frac{l_e^{\prime \prime }}{2l_e^{\prime \prime }+1}}\delta
_{l_e^{\prime },l_e^{\prime \prime }-1}\right] \Phi _{n_e^{\prime
},l_e^{\prime }\rightarrow n_e^{\prime \prime },l_e^{\prime \prime
}}^{n_p,l_p}\quad ,  \label{ec:42}
\end{eqnarray}
\begin{equation}
\Phi _{n^{\prime },l^{\prime }\rightarrow n^{\prime \prime },l^{\prime
\prime }}^{n_p,l_p}=\int_0^\infty R_{n^{\prime \prime },l^{\prime \prime
}}(r)\,R_{n^{\prime },l^{\prime }}(r)\,\Phi _{n_p,l_p}(r)\,r^2\,dr\quad .
\label{ec:39:1}
\end{equation}
\begin{multicols}{2}
An analogous expression to (\ref{ec:42}) holds for ${\cal H}_{\mu
_2\rightarrow \mu _3}^{(h)}$ after the exchange of the subscripts $e$ and $h$%
.

It follows from (\ref{ec:42}) that the quantum numbers of the emitted vibron
are fixed to be $l_p=1$ and $m_p=M_2$. The different states $m_p$ can be
discriminated by selecting adequately the polarization ${\bf e}_j$ of the
exciting light: $M_2=m_p=0$ if ${\bf e}_j\parallel \hat z$ and $M_2=m_p=\pm
1 $ if ${\bf e}_j\parallel (\hat x\pm i\hat y)/\sqrt{2}$. Thus, the HR
selection rules in a spherical QD can be expressed schematically by the
following sequence of states.

\begin{eqnarray}
(L_1=0,M_1=0)\longrightarrow (L_2=1,M_2=0,\pm 1)
\nonumber \\
\stackrel{(l_p=1,m_p=M_2)}{%
\longrightarrow }(L_3=0,M_3=0)\quad ,  \label{seq1}
\end{eqnarray}
which contrasts with that corresponding to Raman scattering \cite
{chamb,eduardo} 
\begin{equation}
(L_1=0,M_1=0)\stackrel{(l_p=0,m_p=0)}{\longrightarrow }(L_2=0,M_2=0)\quad ,
\label{seq2}
\end{equation}
where only $l_p=0$ vibrons can be excited. The comparison between (\ref{seq1}%
) and (\ref{seq2}) makes up clear the complementariness of both light
scattering techniques to study the vibrational spectra of spherical QD's. In
particular, we have shown that hyper-Raman spectroscopy can be used to
investigate the $l_p=1$ vibronic states, not observable in Raman scattering.
In addition, the analysis of the corresponding resonance profiles
(scattering efficiency vs. $\hbar \omega _i+\hbar \omega _j$) can be useful
to reveal $L=1$ exciton states. Of course, the selection rules discussed can
be relaxed when going beyond our simplified treatment of the HR process,
e.g. including valence band mixing. Nevertheless, for $R$ smaller than the
bulk exciton Bohr radius the separation in energy between the {\it hh} and 
{\it lh} levels induced by the confinement should lead to a small amount of 
{\it hh}-{\it lh} admixture. %The amount of admixture decreases as the
%QD radius decreases.

Finally, by inserting Eqs.~(\ref{ec:23}), (\ref{ec:34}), and (\ref{ec:41})
into (\ref{ec:4}), we obtain the following compact expression for the
normalized scattering cross section

\begin{eqnarray}
\frac 1{I_i}\frac{d^{\,2}\sigma ^{ij}}{d\omega _sd\Omega _s}=\sigma _0\left( 
\frac{\bar R}R\right) ^3\sum_{n_p}\,\frac{\eta _s}{\eta _i\eta _j}\,\left( 
\frac{\omega _s}{\omega _j}\right) ^2
\nonumber \\
\left| {\cal M}_{n_p}(\omega _i,\omega
_j,\omega _s;R)\right| ^2\Delta (\omega _s)\quad ,  \label{difer}
\end{eqnarray}
where ${\cal M}_{n_p}$ is the dimensionless amplitude 
\end{multicols}
\begin{equation}  \label{calM}
{\cal M}_{n_p}=E_0^{\,3}\sum_{\mu _1,\mu _2,\mu _3}\frac{f_{\mu
_3}^{*}\left[ {\cal H}_{\mu _2\rightarrow \mu _3}^{(e)}-{\cal H}_{\mu
_2\rightarrow \mu _3}^{(h)}\right] \left[ {\cal F}_{\mu _1\rightarrow \mu
_2}^{(e)}-{\cal F}_{\mu _1\rightarrow \mu _2}^{(h)}\right] f_{\mu _1}}{%
(\hbar \omega _s-E_{\mu _3}+i\Gamma _{\mu _3})(\hbar \omega _i+\hbar \omega
_j-E_{\mu _2}+i\Gamma _{\mu _2})(\hbar \omega _i-E_{\mu _1}+i\Gamma _{\mu
_1})},  \label{Mp}
\end{equation}
\begin{multicols}{2}
and 
\begin{equation}
\sigma _0=2\pi \frac{e^6}{m_0^2\,c^5}\frac{\bar R^2}{(\hbar \omega _i)^2}%
\frac{|{\bf e}_i\cdot {\bf p}_{cv}|^2|{\bf e}_s\cdot {\bf p}_{cv}|^2}{%
m_0^2\,E_0^2}\frac{C_F^2}{E_0^2\bar R}\quad .  \label{I0}
\end{equation}

The expression (\ref{difer}) is suitable for the calculation of the
scattered spectrum as a function of the HR shift, $\omega _s-\omega
_i-\omega _j$. If we are interested in the resonance behavior of the
scattered intensity we can obtain the average HRS efficiency by integrating
Eq.\ (\ref{ec:8:0}) over $\omega _s$, 
\begin{eqnarray}
\left\langle \frac{dS_{HR}^{ij}}{d\Omega _s}\right\rangle =%
\frac{\langle\sigma _0 \rangle}{%
\left\langle V_D\right\rangle }\sum_{n_p}\int \frac{\eta _s}{\eta _i\eta _j}%
\,\left( \frac{\omega _i+\omega _j-\omega _{n_p}(R)}{\omega _j}\right) ^2%
\left( \frac{\bar R}R\right) ^3
\nonumber \\
\left| {\cal M}_{n_p}(\omega _i,\omega
_j,\omega _i+\omega _j-\omega _{n_p}(R);R)\right| ^2\;F(R)dR ,
\label{integra}
\end{eqnarray}
where $\langle \sigma_0 \rangle$ is the average of $\sigma_0$ over 
the QD orientations in the ensemble. 
For typical values of $\bar R$ ($\sim 20$~\AA ) and, $E_0$ ($\sim 10$~meV),
and the parameters of Table~\ref{tab1} a value of 10$^{-4}$~cm\thinspace MW$%
^{-1}$sr$^{-1}$ is estimated for 
$\langle\sigma _0\rangle/\left\langle V_D\right\rangle $.
In the case of an ensemble of QD's the HRS efficiency (\ref{integra}) is
replaced by the average over the QD orientations with the angle 
$\arccos({\bf e}_i\cdot {\bf e}_s)$ fixed.\cite{E}

From now on we focus on the degenerate case, $\omega _i=\omega _j=\omega _l$%
, with 2$\omega _l$ in the region around the excitonic transitions, which is
the usual situation for resonant HRS experiments.

\section{Exciton energies and vibrons in CdSe nanocrystallites}

\label{IV}

In the following we consider for the radial part of the single-particle wave
functions $R_{n,l}(r)$ the solutions of the infinite barrier spherical well
problem 
\begin{equation}
R_{n,l}(r)=\sqrt{\frac 2{R^3}}\frac 1{\left| j_{l+1}(x_n^{(l)})\right| }%
\,j_l\left( x_n^{(l)}\frac rR\right) \quad ,  \label{ec:35:0}
\end{equation}
where $j_l$ is the spherical Bessel function and $x_n^{(l)}$ are the roots
of $j_l(x_n^{(l)})=0$. The single-particle energies are given by 
\begin{equation}
E_{n,l}=\frac{\hbar ^2}{2m_0R^2}\,{x_n^{(l)}}^2\quad .  \label{ec:36:0}
\end{equation}
First, we present in Fig.~\ref{fig:2}(a) the exciton energies, calculated as
explained in Sec.~\ref{II}, for a CdSe quantum dot as a function of the radius.
The material parameters used as input are shown in Table~\ref{tab1}. Solid
and dashed lines correspond to $L=0$ and $L=1$ excitons, respectively. 
%As previously noted the $L=0$ 
%($L=1$) levels will determine the outgoing
%(incoming) resonances in the HR efficiency vs. $2\hbar\omega_l$
%profiles.

According to Ref.~\onlinecite{G} we have taken into account the penetration
of the wave function in the glass matrix by using an effective radius $%
R_{ef}=3.367+0.9078R$ (in \AA ), where $R$ is the nominal QD radius.

Let us focus now the attention on the vibrational modes. Since the active
TO-phonon branch of hexagonal CdSe seems to be flat\cite{tr7}, the parameter 
$\beta _T$ appearing in the isotropic model for the optical vibrations has
been taken as the limit $\beta _T^2\rightarrow 0^{+}$, with a negative bulk
dispersion relation $\omega ^2=\omega _T^2-\beta _T^2q^2$. If one attempts
to solve the vibron model of Ref.~\onlinecite{roca} with $\beta _T=0$, it is
not possible to fullfil all the matching boundary 
conditions at the interface.\cite{tr3} What is expected on physical 
grounds is that the TO component of
the vibron vector displacement, at forbidden frequency for pure TO phonons,
is a vanishing or rapidly decaying function. This can be verified in the
solution for the vibrational amplitude ${\bf u}({\bf r})$ (\thinspace
Eqs.~(35) and (36) of Ref.~\onlinecite{roca}\thinspace ). The TO component
of ${\bf u}$ vanishes for all $r<R$ in the limit $\beta _T^2\rightarrow 0^{+}
$.

Figure~\ref{fig:2}(b) illustrates the allowed frequencies for the $l_p=1$
optical vibrons as a function of the QD radius. Below $\omega _L=213$~cm$%
^{-1}$ it can be seen the confined LO modes with both longitudinal and
transverse components, including a surface mode contribution.\cite{remark}
We can observe some bending in the dispersion around 185~cm$^{-1}$, which is
identified as the Fr\"ohlich mode.\cite{r22} This mode 
%electrostatic potential
arises in the framework of dielectric models, which neglect the effect of
mechanical boundary conditions and consider flat bands for TO and LO bulk
phonons.\cite{remark} The Fr\"ohlich frequency $\omega _F$ is related only
to the dielectric constants of the constituent media: $\omega _F^2=$ $\omega
_T^2(\epsilon _{a0}+2\epsilon _{b\infty })/(\epsilon _{a\infty }+3\epsilon
_{b\infty })$. A strong electrostatic contribution is expected in the
dispersion for vibron frequencies close to $\omega _F$. In Fig.~\ref{fig:3}
we depict the electrostatic potential $\Phi _{n_p,1}(r)$ as a function of $r$
for QD radii $R=11.5$~\AA , $16.2$~\AA , and $21$~\AA . At these radii, $%
\omega _{n_p,1}$ equals $\omega _F$ for $n_p=2,$ 3, and 4. It can be seen
that the electrostatic potential $\Phi _{n_p,1}$ shows an enhancement close
to the interface whenever $\omega _{n_p,1}$ coincides with $\omega _F$.

\section{Scattering intensities}

\label{V}

In this Section the results obtained with Eqs.~(\ref{difer})-(\ref{integra})
for the HRS spectrum and resonance profile in CdSe QD's are shown. The
incoming (outgoing) resonances will be denoted by the combination of the
letter I (O) and the exciton quantum number $N$ of the corresponding
resonant levels. It is important to remind the selection rules already
discussed in Sec.~\ref{II}: The relevant incoming resonances take place at $%
2\hbar \omega _l=E_{N,L=1,M,-1}$ whereas the outgoing ones appear at $\hbar
\omega _s=E_{N,L=0,0,1}$. 
%Thus it is implicitly understood that I (O) resonances 
%are associated to $L=1$ ($L=0$) excitons.

In the next discussion we first analyze the hyper-Raman process for the case
in which %the calculation it has been assumed in Eq.~(\ref{ec:8:0}) that 
all the nanocrystallites in the sample have the same radius $R$. Typical HRS
spectra are shown in Fig.~\ref{fig:4} for different QD radii. The incident
photon energy is such that $\hbar \omega _s=2\hbar \omega _l-\hbar \omega
_{1,1}$ is in resonance with the $N=1,L=0$ excitonic level. We have included
in the calculation the lowest 11 QD excitonic levels ($N=1$ to 11, for each
value $L=0,1$). A lifetime broadening $\Gamma _\mu =5$~meV was assumed for
all excitonic transitions. The spectra are broadened to have a 
full width at half maximum (FWHM) $2\Gamma_{n_p}=4$~cm$^{-1}$.
 It is systematically found that the main peak of the
spectrum corresponds to the creation of the $l_p=1,n_p=1$ vibron, with a
small contribution coming from the $n_p=2,3,4$ vibrons. This is related to
the fact that the matrix element ${\cal H}_{\mu _2\rightarrow \mu _3}^{(e,h)}
$ drops off rapidly as $n_p$ increases. It must be noted that in the spectra
4(a) and 4(b) the $n_p=3$ peak is stronger than the $n_p=2,4$ ones, while
the largest contribution in spectrum 4(c) (aside from the main line) is due
to the $n_p=4$ peak. This difference can be explained in terms of the
electrostatic effects discussed above: If one looks back to the vibron
frequencies as a function of the QD radius in Fig.~\ref{fig:2}(b), it can be
realized that for $R=16$~\AA\ and $18$~\AA {} the vibron frequency $\omega
_{3,1}$ is around the Fr\"ohlich frequency $\omega _F$, and thus the
exciton-vibron interaction is dominant for these modes in the corresponding
spectra 4(a) and 4(b). For $R=21$~\AA , on the other hand, $\omega _{4,1}$
is approximately equal to $\omega _F$ and therefore its potential gives an
enhanced contribution to the exciton-vibron interaction (see also Fig.~\ref
{fig:3}), which is reflected in the spectrum 4(c).

The HR resonance profile (scattering efficiency vs. $2\hbar \omega _l$) for
the $n_p=1,l_p=1$ vibron peak is displayed as a solid line in Fig.~\ref
{fig:5} for QD radius $R=18$~\AA . The most important feature of Fig.~\ref
{fig:5} is that the O1 ($L=0$) resonance takes place at lower energy than
the I1 ($L=1$) resonance, which is a natural consequence from the spacing
between excitonic levels being much larger than the vibron frequencies. It
is worth pointing out that the opposite situation is usually encountered in
bulk semiconductors.\cite{12a} We have also calculated the HRS intensity
taking the Coulomb interaction equal to zero, recovering the free
electron-hole model\cite{chamb} (dashed line), and treating the Coulomb
interaction just in first order perturbation theory (dotted line). It is
apparent the exciton redshift when the electron-hole interaction is
included. Also, when comparing the absolute values of the scattering
intensities for the different approaches we see that the full calculation is
extremely well approximated by the perturbative approach and gives only
slightly larger values than the free electron-hole model (mainly due to the
enhanced oscillator strengths of $L=0$ excitons). Concerning to the HR cross
section the free electron-hole model presents identical lineshape to those
displayed in Fig.~\ref{fig:4}, whenever $2\hbar \omega _l$ is rescaled to
set the equivalent resonance conditions. Thus, we conclude that the
excitonic effects on the HR resonance profile (and also on the HR spectrum)
of quantum dots in the strong confinement regime stand mostly to renormalize
the resonance energies.

Let us finally discuss the effects of the size dispersion of the
crystallites on the HRS spectrum. We have considered an ensemble of QD's
described by a Gaussian distribution function $F(R)$ centered at the mean
radius $\bar R$ and with FWHM equal to 40\%. For a given incoming photon
energy $\hbar \omega _l$, Eqs.~(\ref{ec:7}) determine a set of resonance
radii \{$R_r$\} and their corresponding resonant exciton levels \{$E_r$\}
with lifetime broadenings \{$\Gamma _r$\}. These are marked by arrows in
Fig.~\ref{fig:2}(a), at $2\hbar \omega _l=2.771$~eV. As $R$ departs from $R_r
$, the resonance condition ceases to hold rapidly. The range $\delta R_r$
around the resonance radius $R_r$ which contributes to the integral in Eq. (%
\ref{integra}) is typically of the order of some tens of Angstrom. In such a
small interval, the matrix elements and the vibron frequencies in Eq.~(\ref
{calM}) can be considered as constants. On these grounds we have employed
the following approximation 
\begin{equation}
\int \frac{d^2\sigma (R)}{d\Omega _sd\omega _s}\;F(R)dR\rightarrow \sum_r%
\frac{d^2\sigma (R_r)}{d\Omega _sd\omega _s}\;F(R_r)\delta R_r\quad .
\label{ec:sust}
\end{equation}
An estimation for $\delta R_r$ can be obtained from 
\begin{equation}
\delta R_r=\frac{\pi \Gamma _r}{\left. \frac{dE_r}{dR}\right| _{R_r}}\quad .
\label{ec:deltaR}
\end{equation}

Note that as the HR spectrum evaluated at the resonance radius $R_r$ is
proportional to $\Gamma _r^{-2}$ and $\delta R_r$ is proportional to $\Gamma
_r$, the contribution of the $R_r$-resonant QD's to the average HR spectrum
is proportional to $1/\Gamma _r$.

In Fig.~\ref{fig:6} we show the averaged HR spectrum of the ensemble of CdSe
spherical nanocrystallites with a mean radius of $21$~\AA {}, obtained when $%
2\hbar \omega _l=2.771$~eV. %We have kept the first 12 excitonic levels 
%(for both $L=0$ and $L=1$) in the sums of Eq.~(\ref{calM}). 
This two-photon energy determines a set of incoming and outgoing resonances
with different exciton levels for QD's with different radii $R_r$, which are
listed in Table~\ref{tab2} (for the emission of the $n_p=1,l_p=1$ vibron).
The contributions of the various $R_r$ are also shown in Fig.~\ref{fig:6}.
The resonances due to other exciton levels either are too weak or are
attenuated by the size distribution function. Two important features of the
HR spectrum deserve special attention. The first one is the main peak at
about 208 cm$^{-1}$, which is originated by the emission of the $n_p=1,l_p=1$
vibrons in QD's with different resonance radii. The most important
contribution is that of the outgoing resonance O3 which takes place for $%
R=28.0$~\AA . This resonance appears so strongly because a double resonance
condition conecting the levels $N=3,L=0$ and $N=3,L=1$ is almost fulfilled
(see Fig.~\ref{fig:2}(a)). In addition the resonance I$3$, associated to
dots of radius $R=27.8$~\AA , is very close in energy to O$3$. The second
important feature is the broad structure between 180 and 190 cm$^{-1}$. This
structure is caused by the emission of interface-like vibrons from all
resonant QD's. Its irregular shape is a direct consequence of the vibron
frequency dispersion as a function of QD radius.

Figure~\ref{fig:7} shows the variation of the HR lineshape with small
variations of the incoming photon energy for an ensemble of QD's with a mean
radius of $18$~\AA {}. When $2\hbar \omega _l=2.771$~meV increases
(decreases) by $100$~meV, the radii of resonance get smaller (larger) by
about $1$~\AA\ (see Fig.~\ref{fig:2}(a)). Hence, the principal peak of the
spectrum is shifted to high (low) frequencies, as can be seen in the four
lower curves of the figure. Nevertheless, an opposite effect appears in the
upper spectrum (change from $2\hbar \omega _l=2.771$~eV to $2\hbar \omega
_l=2.971$~eV): In this special situation the higher resonance radii get
closer to the mean radius $\bar R=18$~\AA {} and become dominant in the
spectrum. This effect is equivalent to an increase of the mean radius. It is
important to note that the FWHM of the spectrum at $2\hbar \omega _l=2.771$%
~eV is larger than at $2\hbar \omega _l=2.671$~eV as a result of the
competition between the large contributions of the resonance radii.

Finally, the dependence of the HR lineshape on the mean radius for $2\hbar
\omega _l=2.771$~eV is shown in Fig.~\ref{fig:8}. The main features can be
explained with similar arguments as in the previous figures. For $\bar R=15$
and $16$~\AA {} the lineshapes are practically identical due to the
resonance O1 in the dots of radius $R=18$~\AA . For $\bar R=18$~\AA {} the
maximum is slightly shifted to high frequencies reflecting the influence of
other resonances associated with larger radii. For $\bar R=21$~\AA {} the
main peak is not Lorentzian, showing the contribution of several
equally-strong resonances. And for $\bar R>21$~\AA\ the resonance O3 becomes
dominant.

\section{Conclusions}

\label{VI}

We have developed a formalism to calculate the normalized hyper-Raman
scattering efficiency in spherical semiconductor quantum dots, considering
confined Wannier-Mott excitons as the intermediate states. The
exciton-lattice interaction is assumed to occur via Fr\"ohlich-type
coupling. It has been shown that hyper-Raman spectroscopy can be used to
probe the $l_p=1$ vibrational modes. In addition, each particular mode ($%
m_p=0,\pm 1$) can be selected by properly choosing the polarization of the
incident light. With linearly polarized incident light, only $m_p=0$ vibrons
contribute to the scattering whereas by employing circularly polarized light
the $m_p=\pm 1$ modes are active. The details of the calculations of the
polar optical vibrational modes in CdSe QD's have been discussed and the
eigenfrequencies and electrostatic potentials of the $l_p=1$ modes have been
presented as a function of the nanocrystallite radius. It has been
demonstrated that when their frequencies are close to the Fr\"ohlich
frequency, they undergo an increase in their electrostatic surface character.

The calculations of the hyper-Raman spectra show that the most prominent
peak is due to the emission of $n_p=1,l_p=1$ vibrons. The other
contributions are due to the interface-like $l_p=1$ vibrons. The presence of
surface electronic excitations could increase their role in hyper-Raman
scattering.

The effect of the electron-hole Coulomb interaction has been found to be of
limited importance for the hyper-Raman scattering in the strong confinement
limit relevant to the QD's analyzed here. The absolute values of the HRS
intensity are slightly enhanced by the Coulomb interaction as compared to
the free electron-hole model values. The main excitonic effect seems to be
the renormalization of the resonance energies.

The size dispersion of the nanocrystallites is shown to give rise to a
complex behavior of the main peak position in the HR spectrum as a function
of the laser energy and the details of the distribution over radii. An
additional effect is the distribution of the signal due to interface-like
vibrons over a broad band of frequencies from 180 to 190 cm$^{-1}$.

\acknowledgments

One of us (E. M.-P.) acknowledges J. L. Pe\~na for his hospitality at 
CINVESTAV-IPN (Merida, Mexico), where part of this work was performed. 
We are grateful to Carlos Rodriguez-Castellanos for a critical reading of
the manuscript.

\appendix

\section*{Reduction of the matrix elements}

In this Appendix we show in detail how to calculate matrix elements of the
form 
\begin{equation}  \label{A0}
{\langle
n^{\prime}_e,n_h^{\prime},l^{\prime}_e,l^{\prime}_h,L^{\prime},M^{\prime}%
\vert T_{kq} \vert n_e,n_h,l_e,l_h,L,M \rangle} \quad ,
\end{equation}
where $T_{kq}$ is the $q$th component of a $k$th-order spherical tensor
(single-particle) operator $T_k$, $|n_e,n_h,l_e,l_h,L,M\rangle$ represent an
electron-hole pair state (\,see Eq.~(\ref{ec:12})\,), and $(L,M)$ are the
corresponding total angular momentum quantum numbers.

First of all, we factorize the dependence on $M^{\prime }$, $M$ and $q$ by
applying the Wigner-Eckart theorem: 
\begin{eqnarray}
\ \langle n_e^{\prime },n_h^{\prime },l_e^{\prime },l_h^{\prime },L^{\prime
},M^{\prime }|T_{kq}|n_e,n_h,l_e,l_h,L,M\rangle 
\nonumber \\
=(-1)^{L^{\prime }-M^{\prime}}
\times\left( 
\begin{array}{ccc}
L^{\prime } & k & L \\ 
-M^{\prime } & q & M
\end{array}
\right) 
\sqrt{2L^{\prime }+1}
\nonumber \\
\langle n_e^{\prime },n_h^{\prime },l_e^{\prime
},l_h^{\prime },L^{\prime }\Vert T_k\Vert n_e,n_h,l_e,l_h,L\rangle \quad .
\label{A1}
\end{eqnarray}
The factor $\langle \dots ,L^{\prime }\Vert T_k\Vert \dots ,L\rangle $ does
not depend on $q$ and is called {\it reduced matrix element}. It can be
further simplified by taking into account that $T_{kq}$ is a single-particle
operator. Let us suppose, without loss of generality, that this operator
only acts on the electron coordinates. Then the following reduction formula
can be applied\cite{brink} 
\begin{eqnarray}
\ \langle n_e^{\prime },n_h^{\prime },l_e^{\prime },l_h^{\prime },L^{\prime
}\Vert T_k^{(e)}\Vert n_e,n_h,l_e,l_h,L\rangle =
\nonumber \\
\delta _{n_h^{\prime
},n_h}\,\delta _{l_h^{\prime }l_h}\,(-1)^{k+l_h+L+l_e^{\prime }}\,\left\{ 
\begin{array}{ccc}
L^{\prime } & L & k \\ 
l_e & l_e^{\prime } & l_h
\end{array}
\right\} 
\nonumber \\
\times \sqrt{(2L+1)(2l_e^{\prime }+1)}\langle n_e^{\prime },l_e^{\prime
}\Vert T_k^{(e)}\Vert n_e,l_e\rangle \quad .  \label{A2}
\end{eqnarray}
Finally, by inserting (\ref{A2}) into (\ref{A1}) we achieve the complete
simplification of the matrix element (\ref{A0}).

Let us illustrate this procedure taking as example the matrix element in (%
\ref{ec:29}). As a previous step we must realize that the operator ${\bf e}%
\cdot {\bf r}_e$ can be expressed in terms of spherical irreducible tensors, 
\begin{equation}
{\bf e}\cdot {\bf r}_e=\sum_{q=-1}^1a_qT_{1q}^{(e)}\quad ;\quad
T_{1q}^{(e)}=\sqrt{\frac{4\pi }3}r_eY_{1,q}(\Omega _e),  \label{ec:30}
\end{equation}
with the definitions 
\begin{equation}
a_0=e_z\qquad ,\qquad a_{\pm 1}=\frac{ie_y\mp e_x}{\sqrt{2}}\quad .
\label{ec:31}
\end{equation}
Now, by applying Eqs.~(\ref{A1}) and (\ref{A2}) we obtain 
\end{multicols}
\begin{eqnarray}
\langle n_e^{\prime },n_h^{\prime },l_e^{\prime },l_h^{\prime },L_2,M_2|{\bf %
e}\cdot {\bf r}_e|n_e,n_h,l_e,l_h,L_1,M_1\rangle =\delta _{n_h^{\prime
},n_h}\,\delta _{l_h^{\prime },l_h}(-1)^{1+l_h^{\prime }+L_1+L_2-M_2}
\nonumber \\
\times a_{M_2-M_1}\left( 
\begin{array}{ccc}
L_2 & 1 & L_1 \\ 
-M_2 & M_2-M_1 & M_1
\end{array}
\right) \left\{ 
\begin{array}{ccc}
L_2 & L_1 & 1 \\ 
l_e & l_e^{\prime } & l_h
\end{array}
\right\} \left( 
\begin{array}{ccc}
l_e^{\prime } & 1 & l_e \\ 
0 & 0 & 0
\end{array}
\right) 
\nonumber \\
\times \sqrt{(2L_2+1)(2L_1+1)(2l_e^{\prime }+1)(2l_e+1)}\left( \int
R_{n_e^{\prime },l_e^{\prime }}(r)\,R_{n_e,l_e}(r)\,r^3\,dr\right) \quad .
\label{yay}
\end{eqnarray}
\begin{multicols}{2}
Proceeding in a similar way it can be shown that an expression analogous to (%
\ref{yay}) (with subscripts $e$ and $h$ exchanged everywhere) holds for the
matrix element of the hole operator ${\bf e}\cdot {\bf r}_h$. To obtain the
last expression we have used the following reduced matrix elements 
\begin{equation}
\langle n^{\prime },l^{\prime }\Vert T_1\Vert n,l\rangle =\sqrt{\frac{4\pi }3%
}\left( \int R_{n^{\prime },l^{\prime }}(r)\,R_{n,l}(r)\,r^3\,dr\right)
\langle l^{\prime }\Vert Y_1\Vert l\rangle \quad ,  \label{A7}
\end{equation}
\begin{equation}
\langle l^{\prime }\Vert Y_k\Vert l\rangle =(-1)^{l^{\prime }}\sqrt{\frac{%
(2l+1)(2k+1)}{4\pi }}\left( 
\begin{array}{ccc}
l^{\prime } & k & l \\ 
0 & 0 & 0
\end{array}
\right) \quad .  \label{A4}
\end{equation}

Moreover, if we particularize (\ref{yay}) to the case $L_1=M_1=0$ ($l_e=l_h$%
) and evaluate the $3j$ and $6j$ symbols we get the result shown in Eq.~(\ref
{ec:32}).

Applying the same method the following expression is obtained for the matrix
elements of the electron-lattice interaction
\end{multicols} 
\begin{eqnarray}
\langle n_e^{\prime \prime }n_h^{\prime \prime }l_e^{\prime \prime
}l_h^{\prime \prime }L_3M_3|e\hat \varphi _F({\bf r}_e)|n_e^{\prime
}n_h^{\prime }l_e^{\prime }l_h^{\prime }L_2M_2\rangle =\delta _{n_h^{\prime
},n_h^{\prime \prime }}\delta _{l_h^{\prime },l_h^{\prime \prime
}}(-1)^{L_2+L_3+l_p+l_h^{\prime }-M_2}
\frac{C_F}{\sqrt{R}}\sqrt{(2L_2+1)(2L_3+1)}  
\nonumber \\
\sum_{n_p,l_p,m_p} \sqrt{%
(2l_e^{\prime }+1)(2l_e^{\prime \prime }+1)(2l_p+1)/4\pi } 
\left( 
\begin{array}{ccc}
L_2 & l_p & L_3 \\ 
-M_2 & m_p & M_3
\end{array}
\right) \left( 
\begin{array}{ccc}
l_e^{\prime } & l_p & l_e^{\prime \prime } \\ 
0 & 0 & 0
\end{array}
\right) \left\{ 
\begin{array}{ccc}
L_2 & L_3 & l_p \\ 
l_e^{\prime \prime } & l_e^{\prime } & l_h^{\prime }
\end{array}
\right\} \Phi _{n_e^{\prime \prime },l_e^{\prime \prime }\rightarrow
n_e^{\prime },l_e^{\prime }}^{n_p,l_p}\quad .  \label{ec:39}
\end{eqnarray}
\begin{multicols}{2}
In this expression it is implicit that only the creation part of $\hat 
\varphi _F$ is acting on the ket. Replacing $L_2=1,L_3=M_3=0$, the Eq.~(\ref
{A4}) is reduced to Eq.~(\ref{ec:42}) and $\Phi _{n_e^{\prime \prime
},l_e^{\prime \prime }\rightarrow n_e^{\prime },l_e^{\prime }}^{n_p,l_p}$ is
defined in Eq.~(\ref{ec:39:1}).

%\newpage
\end{multicols}

\begin{figure}
\caption{Feynman diagrams which give the main contribution to the resonant
hyper-Raman scattering.}
\label{fig:1}
\end{figure}

\begin{figure}
\caption{(a) Energy levels of $N=1,...7$ and $L=0$ (solid lines) and $L=1$
(dashed lines) excitons in a spherical CdSe nanocrystallite as a function of
its radius. Dashed line I indicates the two-photon energy $2\hbar \omega
_l=2.771$~eV and solid line II corresponds to the scattered photon energy $%
\hbar \omega _s=2\hbar \omega _l-\hbar\omega_{1,1}=2.745$~eV. The solid
(dashed) arrows indicate the radii of the corresponding outgoing (incoming)
resonances. Line III represents a Gaussian distribution over QD radii
centered at $\bar R=18$~\AA\ and with FWHM equal to 40\% (see
text). (b) Frequency of the $l_p=1$ optical vibrons of a CdSe
nanocrystallite as a function of its radius.}
\label{fig:2}
\end{figure}
\begin{figure}
\caption{Plot of the electrostatic potentials associated to the first
optical vibrons for different crystallite radii. Solid line: $n_p=1$; dashed
line: $n_p=2$; dotted line: $n_p=3$; dot-dashed line: $n_p=4$. The equation $%
\omega_{n_p,1}(R)=\omega_F$ with $n_p=2$, $3$, and $4$ is fulfilled for $%
R=11.5$~\AA{}, $R=16.2$~\AA{}, and $R=21$~\AA{}, respectively. We see that
the electrostatic potential $\Phi_{n_p,1}$ increases at the interface
whenever $\omega_{n_p,1}$ coincides with the Fr\"{o}hlich frequency $%
\omega_F $. }
\label{fig:3}
\end{figure}
\begin{figure}
\caption{Hyper-Raman spectra of CdSe nanocrystallites embedded in glass for
different radii and two photon energies: a) $R=16$~\AA {} and $2\hbar \omega _l=2.979$~eV . b) $R=18$%
~\AA {} and $2\hbar \omega _l=2.775$~eV. c) $R=21$~\AA{} and $2\hbar \omega
_l=2.559$~eV.}
\label{fig:4}
\end{figure}
\begin{figure}
\caption{Hyper-Raman intensity calculated for CdSe nanocrystallites with
radius $R=18$~\AA {}. The resonances are denoted by the labels I$N$ or O$N$,
where I (O) means incoming (outgoing) resonance with the exciton level $N$.
Three approaches have been used: Solid line: Full matrix diagonalization
including exciton effects. Dotted line: first order perturbation theory for
the energy. Dashed line: free electron-hole model.}
\label{fig:5}
\end{figure}
\begin{figure}
\caption{Hyper-Raman spectrum for $2\hbar \omega _l=2.771$~eV of an ensemble
of CdSe QD's with mean radius 21~\AA {} and a 40\% size dispersion. The
contributions of the QD's with different resonance radii are also shown. }
\label{fig:6}
\end{figure}
\begin{figure}
\caption{Hyper-Raman spectra for an ensemble of QD's with a mean radius $18$%
~\AA {} and a 40\% size dispersion for different values of the laser energy. The 
inset illustrates the dependence of the maximum position and FWHM on the two 
photon energy $2\hbar\omega_l$. The lines are a guide to the eyes.}
\label{fig:7}
\end{figure}
\begin{figure}
\caption{ Dependence of the HR lineshape with the mean radius $\bar R$ at $%
2\hbar \omega_l=2.771$~eV. The size dispersion is 40\% of the mean radius
in all the spectra. }
\label{fig:8}
\end{figure}

\newpage

\narrowtext
\begin{table}
\caption{ Values of the material parameters used for the numerical
calculations. The same value $\Gamma_{\mu}$ ($\Gamma_{n_p}$) has been
assigned to all exciton (vibron) states. }
\label{tab1}
\begin{tabular}{lc}
Parameter & Value \\ \hline
$m_e/m_0$ & 0.12\tablenotemark[1] \\ 
$m_h/m_0$ & 0.45\tablenotemark[1] \\ 
$E_g$ (eV) & 1.865\tablenotemark[2] \\ 
$\omega_L$ (cm$^{-1}$) & 213.1\tablenotemark[2] \\ 
$\omega_T$ (cm$^{-1}$) & 165.2\tablenotemark[3] \\ 
$\beta_L$ ($10^3$ms$^{-1}$) & 2.969\tablenotemark[2] \\ 
$\beta_T$ ($10^3$ms$^{-1}$) & $0.002$ \\ 
$\epsilon_{a0}$ (CdSe) & 9.53\tablenotemark[4] \\ 
$\epsilon_{a\infty}$ (CdSe) & 5.72\tablenotemark[5] \\ 
$\epsilon_{b\infty}$ (Glass) & 4.64\tablenotemark[6] \\ 
$\Gamma_{\mu}$ (meV) & 5 \\ 
$\Gamma_{n_p}$ (cm$^{-1}$) & 2 \\ 
$P^2$ (eV) & 20\tablenotemark[7] \\ 
\end{tabular}
%\par
\tablenotetext[1]{Ref.~\onlinecite{landolt}.}
\tablenotetext[2]{Ref.~\onlinecite{G}.}
\tablenotetext[3]{Ref.~\onlinecite{tr7}.}
\tablenotetext[4]{Ref.~\onlinecite{alonso}.}
\tablenotetext[5]{Calculated from the
Lydanne-Sachs-Teller relation.}
\tablenotetext[6]{Ref.~\onlinecite{31}.}
\tablenotetext[7]{Ref.~\onlinecite{hermann}.
 $P^2=2|{\bf p}_{cv}|^2/m_0$}.
\end{table}

\begin{table}
\caption{List of resonance radii and corresponding resonant exciton levels
(see text and Fig.~6). The incoming (outgoing) resonances are labeled by 
I$N$ (O$N$)}
\label{tab2}
\begin{tabular}{c|cccccc}
$R$ (\AA) & 18.0 & 20.3 & 26.0 & 27.8 & 28.0 & 30.0 \\ 
$(N,L)$ & (1,0) & (1,1) & (2,1) & (3,1) & (3,0) & (4,1)\\
    -   &   O1  &  I1   &   I2  &   I3  &  O3   &  I4 
\end{tabular}
\end{table}

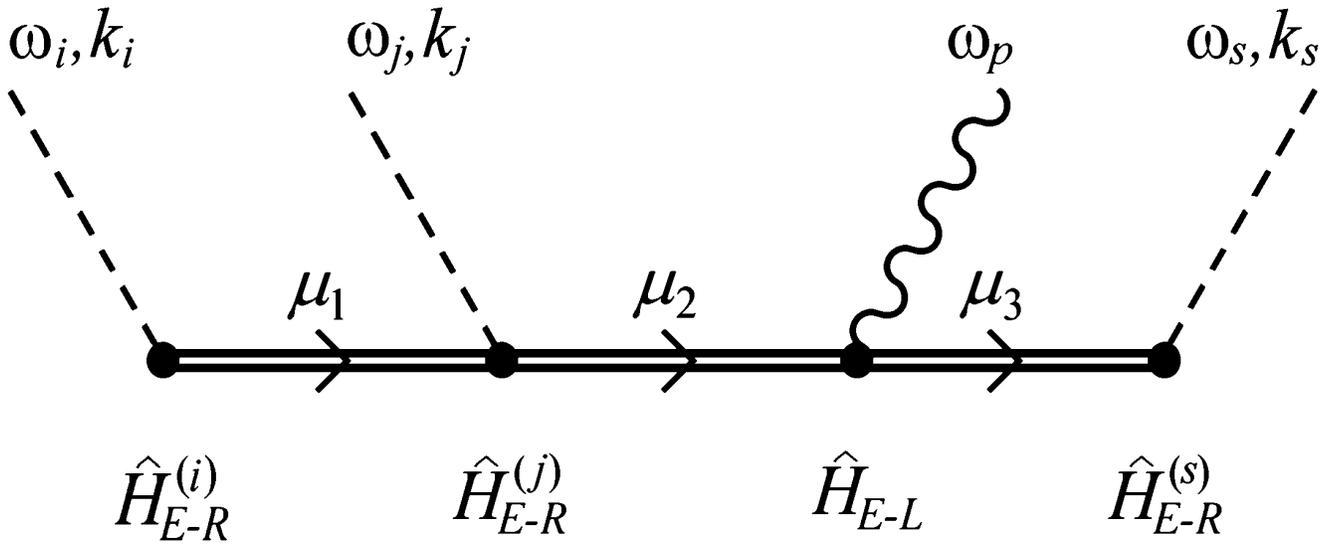
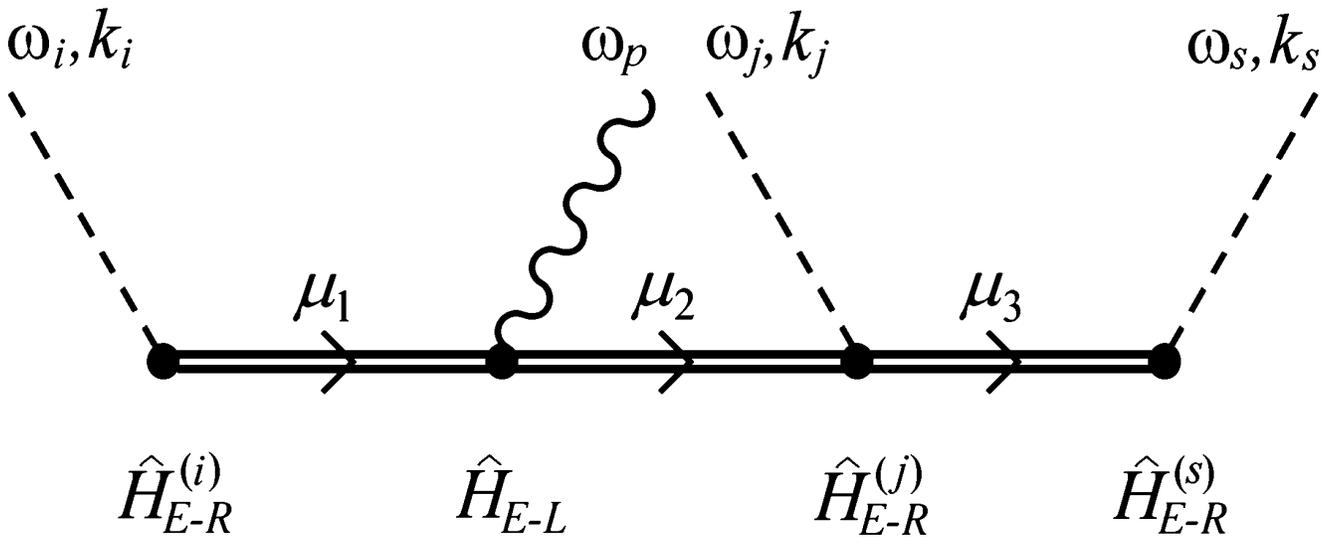
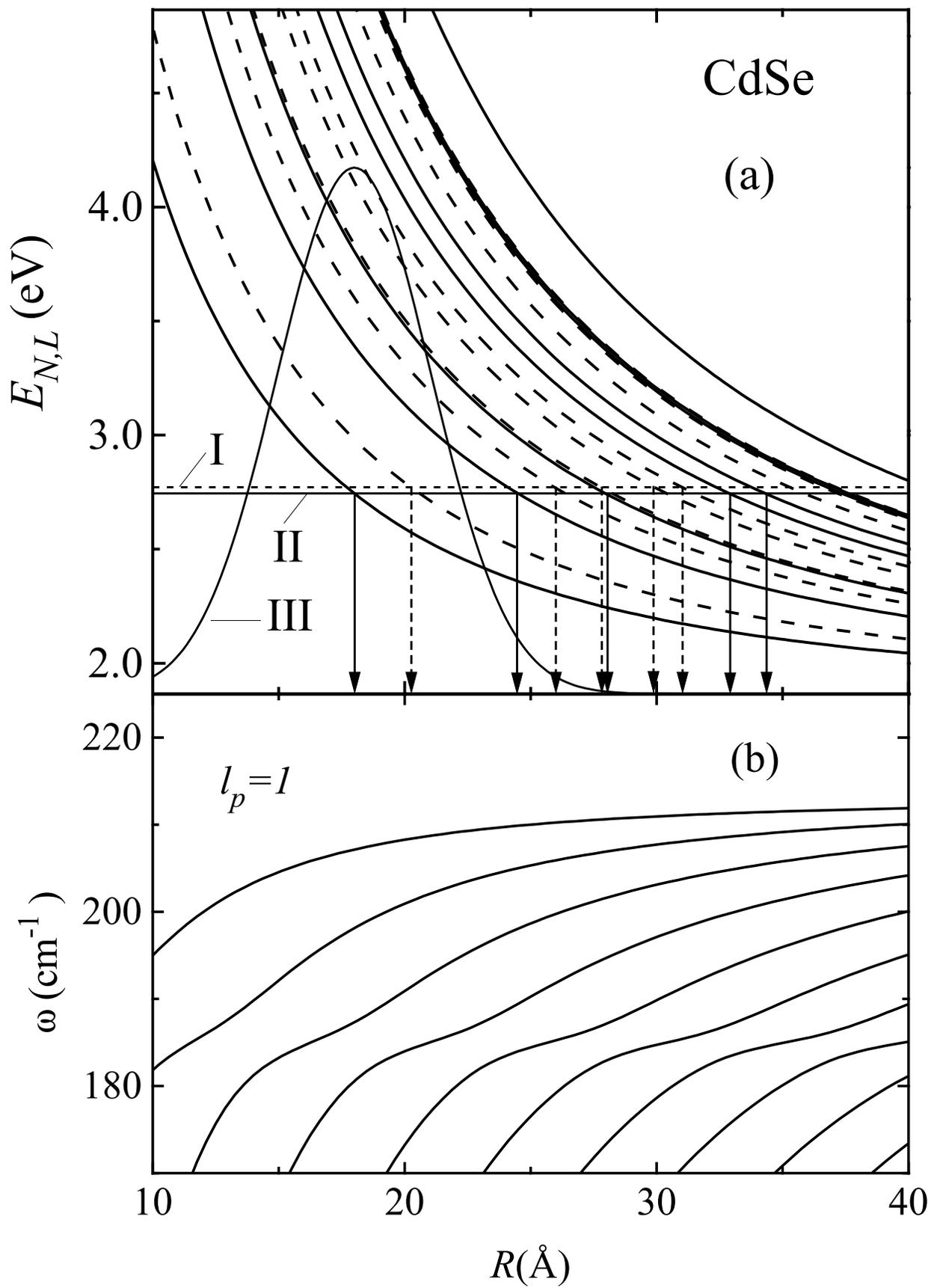
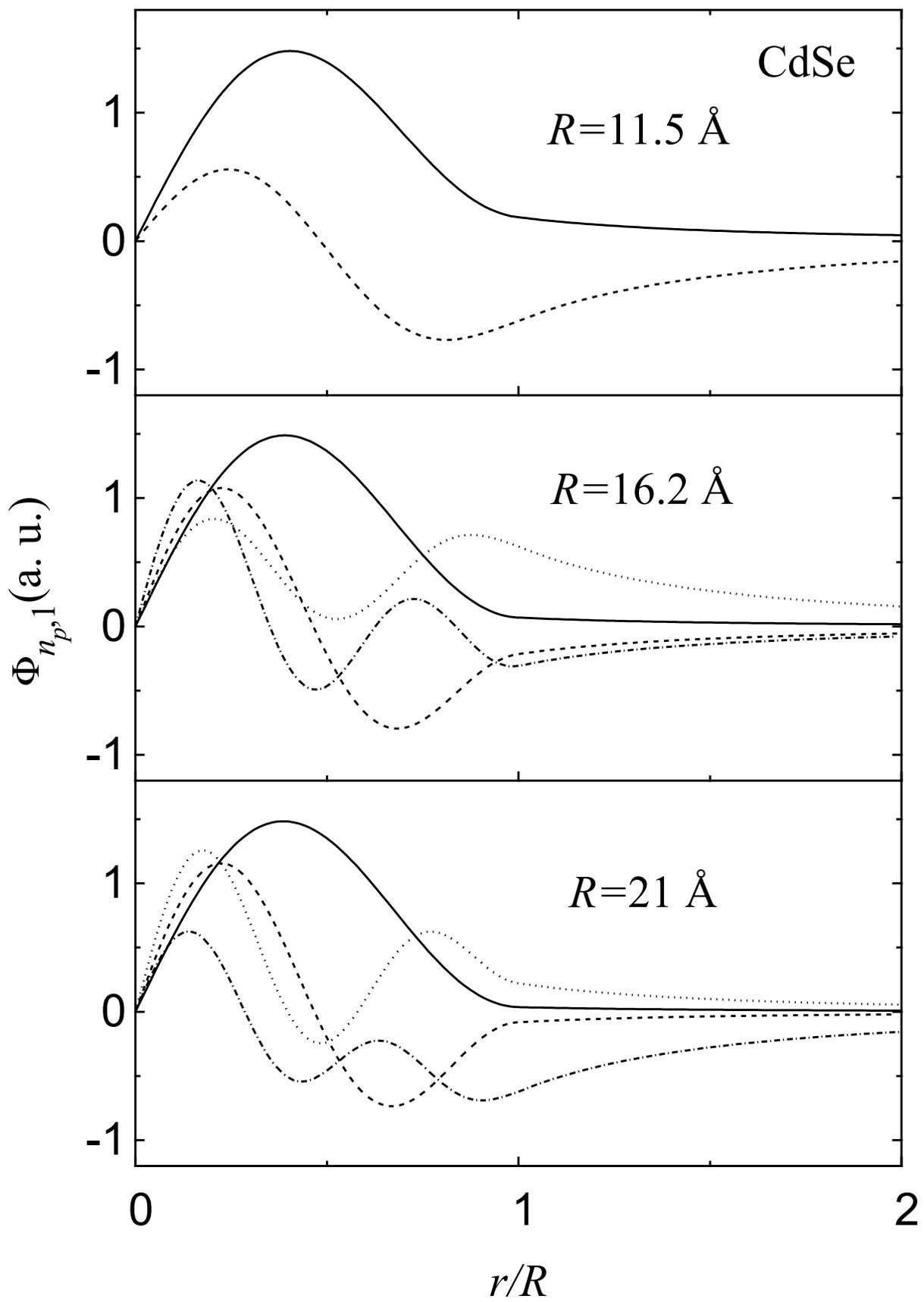
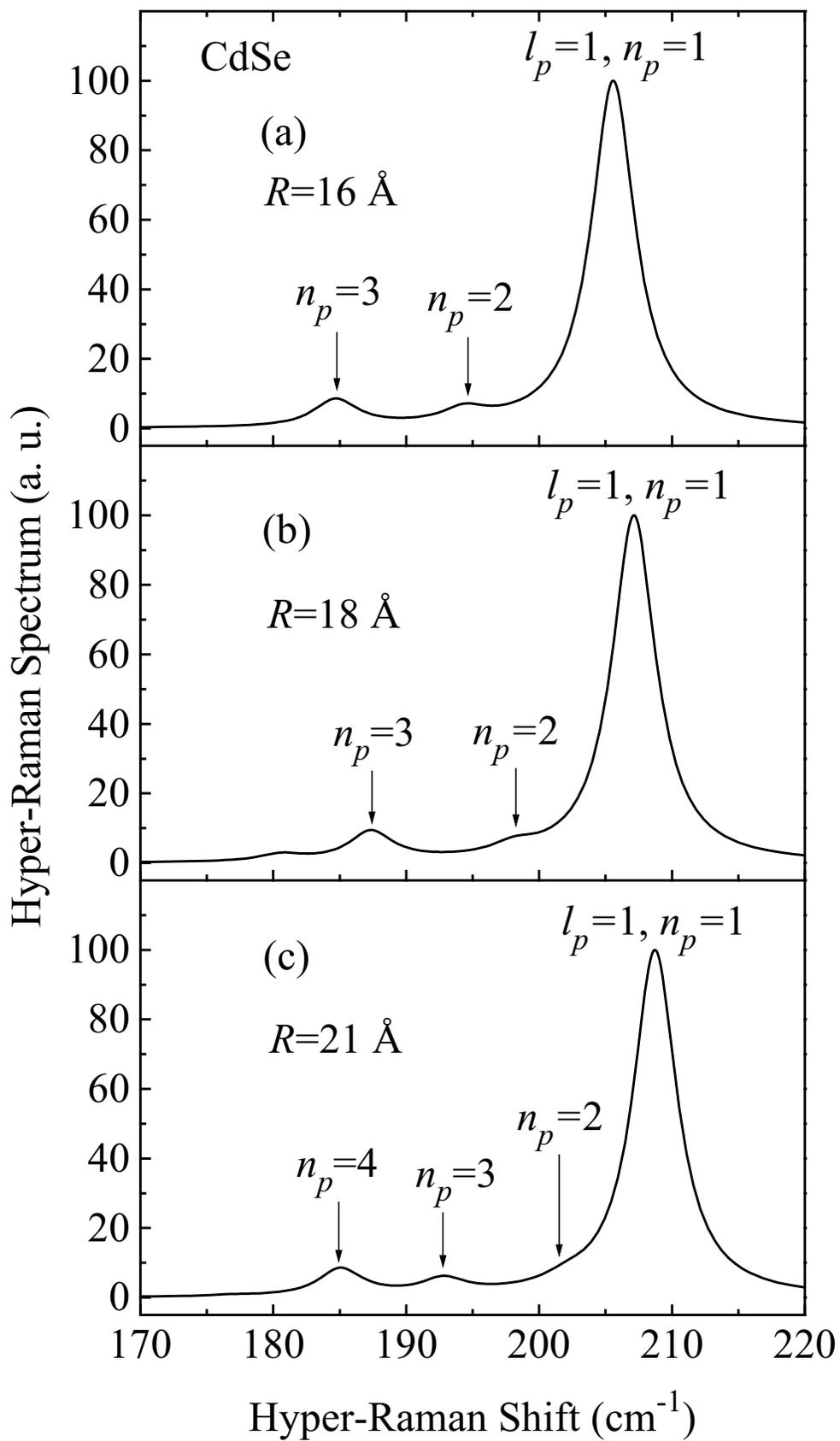
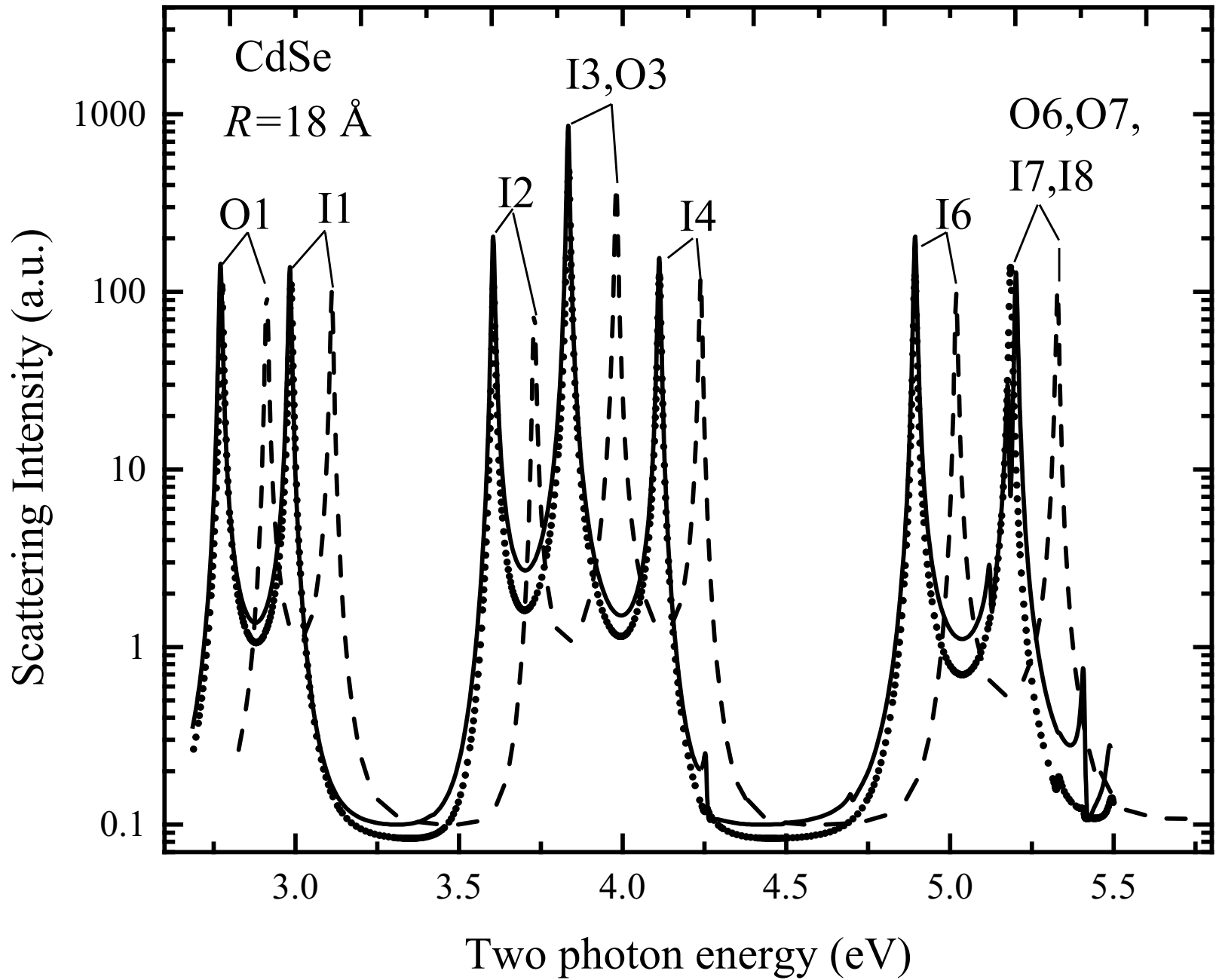
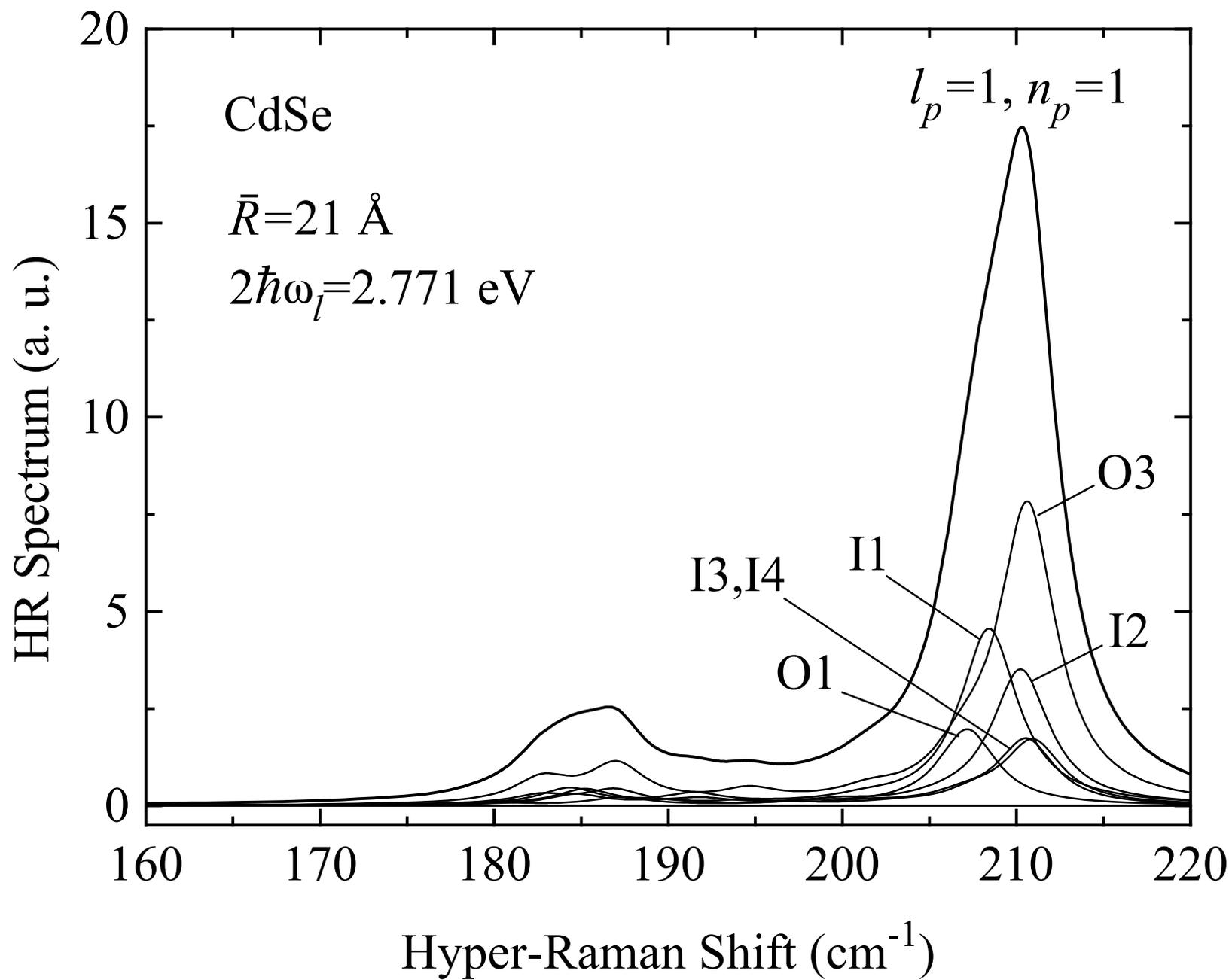
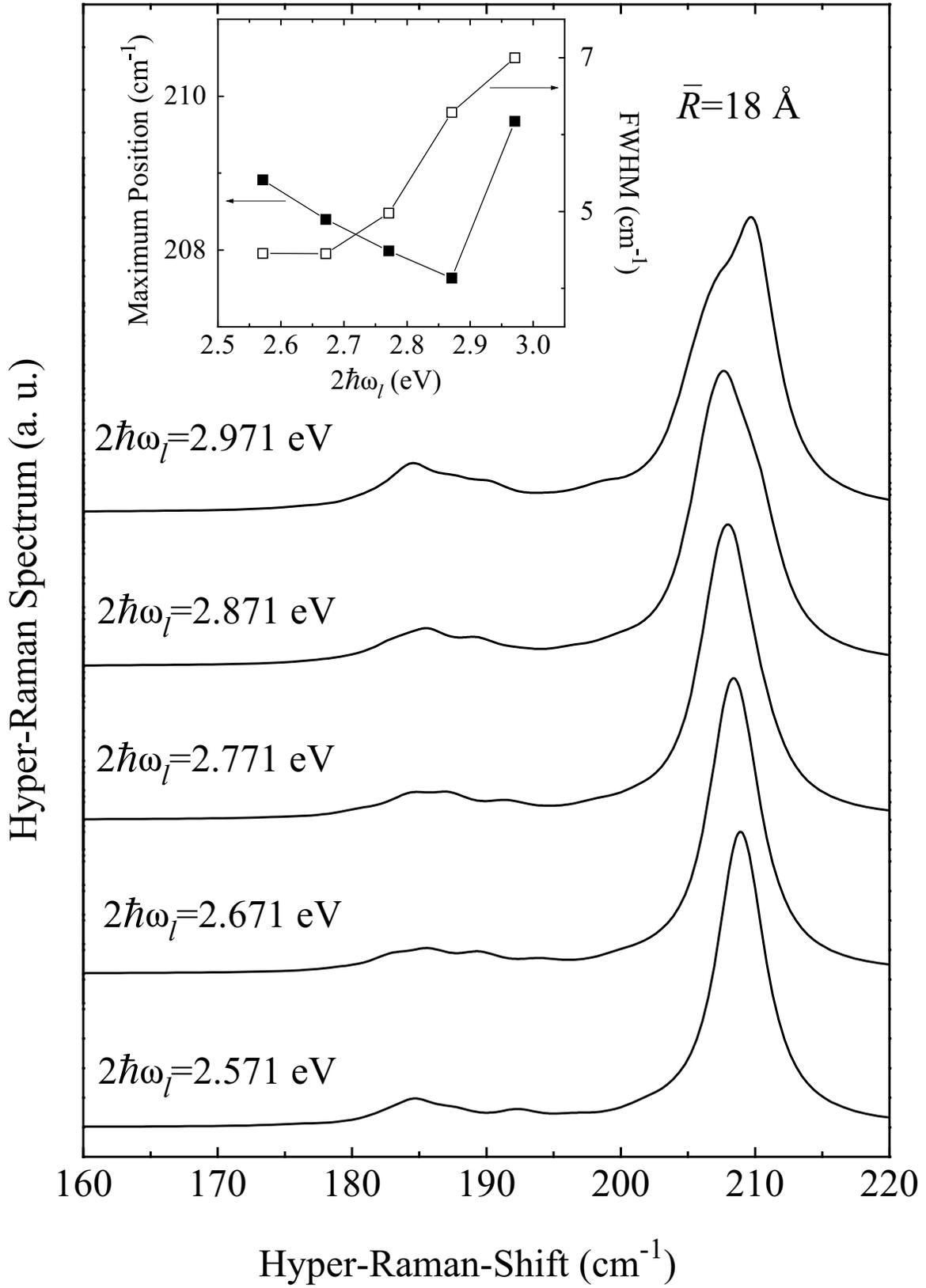
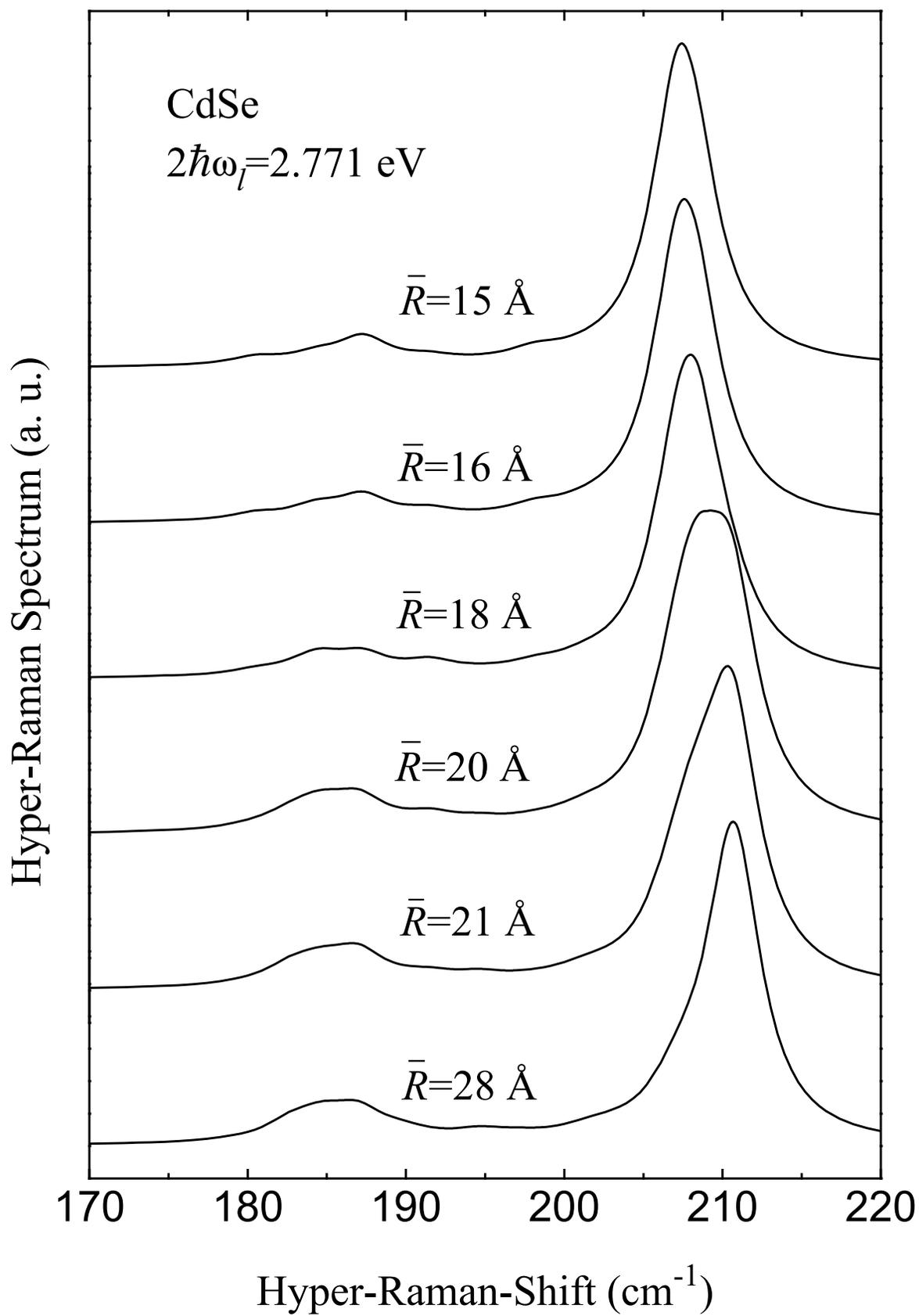
\end{document}